\documentclass[aps,pre,twocolumn,superscriptaddress,showpacs]{revtex4-1}
\usepackage{amsmath,amssymb,graphicx,color}
\usepackage{bm}
\usepackage{bbm}
\def\Tr{\mbox{Tr}}
\begin{document}
\title{Aspects of Work}
\author{ Peter Talkner}
\affiliation{Institut f\"{u}r Physik, Universit\"{a}t Augsburg, Universit\"{a}tsstra{\ss}e 1, D-86135 Augsburg, Germany}
\affiliation{Institute of Physics, University of Silesia, 40007 Katowice, Poland}
\author{Peter H\"anggi}
\affiliation{Institut f\"{u}r Physik, Universit\"{a}t Augsburg, Universit\"{a}tsstra{\ss}e 1, D-86135 Augsburg, Germany}
\affiliation{Nanosystems Initiative Munich, Schellingstr, 4, D-80799 M\"unchen, Germany} 
\date{\today}
\begin{abstract}
Various approaches of defining and determining work performed on a quantum system are compared. Any operational definition of work, however, must allow for two facts, first, that work characterizes a process rather than an instantaneous state of a system, and, second, that quantum systems are sensitive to the interactions with a measurement apparatus. 
We compare different measurement scenarios on the basis of the resulting post-measurement states and the according probabilities for finding a particular work value. In particular, we analyze a recently proposed work-meter for the case of a Gaussian pointer state and compare it 
with the results obtained by two projective and, alternatively, two Gaussian measurements. 
In the limit of a strong effective measurement strength the work distribution of projective two energy measurements can be recovered. In the opposite limit the average of work becomes independent of any measurement. Yet the fluctuations about this value diverge.  The performance of the work-meter is illustrated by the example of a spin in a suddenly changing magnetic field.         
\end{abstract}
\maketitle
\section{Introduction}\label{intro}
Work belongs to the most basic notions of classical mechanics and also presents one of the cornerstones of thermodynamics. With the recent experimental progress in the fields of cold atoms\cite{BDZ}, nano-mechanics, and opto-mechanics, \cite{KV,AKM} to name just a few, there is an urgent need of a theoretical foundation of what work means in quantum mechanics and how it can be defined in an operational way. 

One reason why work does not simply fall into the category of observables like position, linear and angular momentum, and energy, comes from the fact that it is meant to characterize a process rather than an instantaneous state of a system \cite{TLH}. 
Even in the simple case of a thermally isolated system the determination of work requires an interference of the system with a measurement device at two times. 
For a classical system, the interaction between  the system and the measurement device can be made arbitrarily weak and, hence, the back-action can be arbitrarily small, without implying a principle limitation of the precision of the measurement. 
However, for a quantum systems, the back-action of the measurement device  
modifies the state of the system; it consequently influences the outcome of a subsequent contact, therefore influencing the final value of the work \cite{TMH}.

The paper is organized as follows. In Section \ref{classical} we survey  various definitions of work that classically are equivalent to each other. We review in more detail the operational definition in terms of two projective energy measurements in Section \ref{tema}. In Section \ref{twoenergy} we discuss a Gaussian energy measurement scheme which then is employed in a two Gaussian energy measurement device.
A recently proposed approach to determine work by a single, necessarily generalized measurement \cite{RCP,dCRP} is discussed in Section\ref{meter} for a Gaussian pointer.  
A work-meter that functions in that way operates with a precision determined by a parameter combining the strength of the measurement and the covariance matrix of the Gaussian initial pointer state. Implementing a  high precision the result of two projective energy measurements is recovered while the use of low precision leads to broad distributions yet with a finite average. A spin in a suddenly changing magnetic field serves as an example in Section \ref{example}. Section \ref{con} concludes the paper.    
\section{Classically equivalent definitions of work and their  nonequivalent quantum counterparts} \label{classical}
Restricting ourselves to thermally isolated systems we begin with the discussion of work in classical systems. When a classical system stays at a phase space point $z$ its  energy is determined by the value of a  conveniently chosen Hamiltonian $H(z,\lambda)$ \cite{gauge} which also governs the dynamics of the system. A change of the parameter $\lambda$ according to a protocol $\Lambda = \{ \lambda(t)|0\leq t\leq \tau \}$ alters the energy and hence work will be applied to or taken from the system. We shall adopt the inclusive definition of work \cite{J07} according to which  
the work is given by the energy-difference at the end and the beginning of the force protocol, 
\begin{equation}
w=H\big (Z(\tau,z_0),\lambda(\tau)\big)-H\big (z_0,\lambda(0)\big )
\label{wHHcl}
\end{equation}
with $Z(t,z_0)$ being the phase space point that evolves from $Z(0,z_0)=z_0$ according to the Hamiltonian dynamics, $\dot{Z} = \{ H(Z,\lambda(t)),Z\}_P$
where $\{ , \}_P$ denotes the Poisson bracket. As already mentioned, within the time-range $[0,\tau]$ the parameter $\lambda(t)$ changes according to a prescribed protocol  $\Lambda$.
For  classical systems an equivalent definition of work is based on the fact that the energy difference can be expressed as an integral of the total time-derivative of the Hamiltonian, which coincides with the respective partial derivative yielding the work as integral of the supplied power, i.e., 
\begin{equation}
w=\int dt \frac{\partial H(Z(t,z_0,\lambda(t))}{\partial \lambda(t)}  \dot{\lambda}(t)\:, 
\label{wpower}
\end{equation}
where $\dot{\lambda}(t)$ denotes the time-derivative of the force parameter $\lambda(t)$.

Any adaptation of the  two work definitions (\ref{wHHcl}) and (\ref{wpower}) to quantum systems must take into account the back-actions of those measurements that have to be performed in order to determine the work. It is this inherently quantum-mechanical requirement that renders the two classically equivalent definitions (\ref{wHHcl}) and (\ref{wpower}) inequivalent \cite{TLH,VWT15}. Based on the energy difference definition (\ref{wHHcl}) work can be determined by the {\it two energy measurement approach} \cite{Kurchan,Tasaki,TLH} which employs two projective measurements of energy at the end and the beginning of the force protocol. Due to the projective nature of these measurements, two energy-eigenvalues $e_m(\tau)$ and $e_n(0)$ are obtained. From these eigenvalues the work in a single realization of the protocol is determined as the difference $w =e_m(\tau)-e_n(0)$. With this definition the statistics of work complies with the fluctuation relations of Jarzynski \cite{Jarzynski97} and Crooks \cite{Crooks} for systems initially staying in a canonical initial state \cite{JC}. 

A quite different picture emerges if one tries to implement the power based approach \cite{VWT15}. This, in principle, requires a continuous observation of the system in such a way that the time-evolution of the power can be inferred. Even if this monitoring is not performed by means of projective measurements but in a weaker way, a continuous observation will freeze the dynamics of the system due to the quantum Zeno effect \cite{VWT15}. For relatively slowly varying protocols one might try to circumvent the Zeno effect by restricting the observations to a discrete set of times with a suitably chosen strength of observation. Yet the resulting statistics of work turn out to be essentially determined by the number and strength of the observations and only to a lesser extent by the force protocol. There are hardly any relations between the statistics obtained by the two energy measurement and the power approach. In particular, the fluctuation relations of Crooks and Jarzynski are only satisfied for the power approach if the Hamiltonians at all different parameter values commute with each other \cite{VWT15}. This is not in contradiction with the fact that measurements during the force protocol leave these fluctuation relations unchanged as long as the work is determined by two projective measurements at the beginning and the end of the force protocol \cite{CTH10,WVTCH,VWT15}.        
  
We finally note that in various attempts to formulate a ``quantum thermodynamics'' \cite{quth} work is understood in a way such that its mean-value coincides with the difference of the mean-values of the energies at the end and the  beginning of the force-protocol, however, without taking into account the back-action of the first onto the second measurement. In order to distinguish this notion of a work-like quantity from operational definitions that include the back-action of measurements we call it ``untouched work'' \cite{untouched}. In this sense, work is always untouched in classical mechanics.

The defining property of untouched work hence postulates that its average 
$\prec w \succ$ can be expressed as 
\begin{equation}
\prec w \succ = \Tr U^\dagger_\Lambda H\big (\lambda(\tau)\big ) U_\Lambda \rho(0) - \Tr H\big (\lambda(0)\big ) \rho(0)\:,
\label{varw}
\end{equation} 
where the density matrix $\rho(0)$ describes the initial state and 
$U_\Lambda\equiv U_{\tau,0}$ denotes the unitary time-evolution of the system over the extension of the force protocol, following as the solution of the Schr\"odinger equation
\begin{equation}
i \hbar \partial U_{t,s} /\partial t = H\big (\lambda(t)\big ) U_{t,s}\:, \quad U_{s,s} = \mathbbm{1}\:,
\label{SE}
\end{equation}
where $\mathbbm{1}$ denotes the unit operator on the Hilbert space of the system.

For later use, we introduce the spectral resolution of the time-dependent Hamiltonian $H\big (\lambda(t) \big )$ in terms of its eigenvalues $e_n(t)$ and corresponding projection operators $\Pi_n(t)$ satisfying $\Pi_n(t) \Pi_m(t) = \delta_{m,n} \Pi_m(t)$, $\Pi^\dagger_n(t) = \Pi_n(t)$ and $\sum_n \Pi_n(t) = \mathbbm{1}$.  The Hamiltonian can then be represented as
\begin{equation}
H\big (\lambda(t) \big ) = \sum_n e_n(t) \Pi_n(t)\:.
\label{Hspec}
\end{equation}
We will make frequent use of this representation in the sequel for the initial and final times $t=0,\tau$, respectively.

A way to obtain the average untouched work would be to prepare two copies of an ensemble representing the initial state of the system. The first ensemble is then used to determine its average energy in the standard way by projective measurements of the initial energy. The ensemble which is generated by this measurement need not coincide with the one prior to the measurement and is discarded after the measurement. The force protocol is applied on the members of the second ensemble and upon its completion the energy is measured and its average calculated. 
The difference of the two average energies coincides with the average of the untouched energy. 
It only agrees with the average work obtained by the above mentioned two energy measurement approach if the discarded  ensemble was identical with the original one. This is the case for an initial state that is stationary with respect to the initial Hamiltonian. 
The given prescription though is incomplete in that it only determines an average value but does not allow to determine fluctuations in a meaningful way \cite{fluctuations}. 
Attempts to define the work as an observable in terms of the difference of the initial and final Hamiltonians in the Heisenberg-picture \cite{AN,EN} disregard the inherent process-dependence of work \cite{TLH,CHT,CHTe,HT}. Also for a canonical initial state, the Jarzynski and Crooks relations do not hold with this definition apart from the trivial, yet untypical case of commuting Hamiltonians, $[H\big (\lambda(t)\big ),H\big (\lambda(t')\big)]=0$ for all $0\leq t,t' \leq \tau$.
  
In a recent effort,  Allahverdyan \cite{Allah} suggested a characterization of untouched work as a fluctuating quantity which indeed yields (\ref{varw}) for the average. 
The statistics of this hypothetical object  is determined in \cite{Allah} by a pdf-like function $q(w) = \sum_{m,n} \delta(w-e_m(\tau)+e_n(0)) p^{\text{TMH}}_{m,n}$ 
The weight $p^{\text{TMH}}_{m,n}$ is a quasi-probability for the eigenvalues of the two Hamiltonians  $H\big (\lambda(0)\big )$ and $H\big (\lambda(\tau)\big )$ based on the Terletsky-Margenau-Hill distribution which is given by $p^{\text{TMH}}_{m,n}= \Tr \rho(0) \{U^\dagger_\Lambda \Pi_m(\tau) U_\Lambda,\Pi_n(0) \}/2 $ with $\{A,B\} = AB +BA$ denoting the anti-commutator of two operators $A$ and $B$. The fact that, as a quasi-probability, the Terletsky-Margenau-Hill distribution assumes negative values for non-commuting Hamiltonians  may also translate to $q(w)$ which hence is not a probability density function and consequently fails to characterize a proper random variable.

\section{Work from two projective energy measurements}\label{tema}
The joint probability $p_\Lambda(m,n)$ of observing the system at an eigenvalue $e_n(0)$ of the Hamiltonian $H\big (\lambda(0)\big )$ by projective measurements at the beginning of the force protocol $\Lambda$ and the eigenvalue $e_m(\tau)$ of the Hamiltonian $H\big (\lambda(\tau)\big )$ at its end 
is given by \cite{CHT,CHTe}
\begin{equation}
p_\Lambda(m,n) = \Tr \Pi_m(\tau) U_\Lambda \Pi_n(0) \rho(0) \Pi_n(0) U^\dagger_\Lambda\:,
\label{pmn}
\end{equation}
where the projection operators $\Pi_n(t)$, at times $t=0,\tau$, are defined by the spectral representations of the respective Hamiltonians, see (\ref{Hspec}). It is important to note that due to the first energy measurement the initial density matrix $\rho(0)$ is projected onto the eigenbasis of the initial Hamiltonian $H\big (\lambda(0)\big )$. In other words, all non-diagonal elements of the initial density matrix in the energy eigenbasis are erased by the first measurement. This will influence the subsequent time evolution unless the initial density matrix and Hamiltonian commute with each other. Moreover, this projection excludes the use of correlations between energy eigenstates as a ``resource'' to extract work out of a system by a unitary process, as it has been recently suggested \cite{Alicki,Perarnau}.  

The probability density function (pdf) $p^{\text{pem}}_\Lambda(w)$ of finding the work $w$ by means of two projective energy measurements (pem) can be expressed in terms of the joint probability (\ref{pmn}) as \cite{CHT}
\begin{equation}
p^{\rm{pem}}_\Lambda(w) = \sum_{m,n} \delta \big( w - e_m(\tau) + e_n(0)\big) p_\Lambda(m,n)\:.
\label{pw}
\end{equation}

In a realization of the force-protocol for which the work $w$ was obtained
the state $\rho_w(\tau^+)$, immediately after the second energy measurement, is proportional to a   linear, positive map $\Phi_w$ acting on the initial density matrix $\rho(0)$ with a proportionality constant inversely proportional to the work pdf $p^{\rm{pem}}_\Lambda(w)$. Hence, the post-measurement state can be written as
\begin{equation}
\rho_w(\tau^+) = \Phi_w(\rho(0)) / p^{\rm{pem}}_\Lambda(w)\:,
\label{rPhip}
\end{equation}
where the so-called operation $\Phi_w$ \cite{Kraus} is given by
\begin{equation}
\begin{split}
\Phi_w(\rho)& = \sum_{m,n} \delta(w-e_m(\tau) +e_n(0))\\
&\quad \times \Pi_m(\tau) U_\Lambda \Pi_n(0) \rho \Pi_n(0) U^\dagger_\Lambda \Pi_m(\tau)\:.
\end{split}
\label{rPhir}
\end{equation}
If the energies are measured but the work is not registered the system is found in the so-called non-selective post-measurement state $\rho(\tau^+)$, given by
\begin{equation}
\begin{split}
\rho(\tau^+) &= \int dw \rho_w(\tau^+) p^{\text{pem}}_\Lambda(w)\\
&= \int dw\: \Phi_w(\rho(0))
\label{rhous}
\end{split}
\end{equation}
leading with (\ref{rPhir}) to
\begin{equation}
\rho(\tau^+) = \sum_{m,n} \Pi_m(\tau) U_\Lambda \Pi_n(0) \rho(0) \Pi_n(0) U^\dagger_\Lambda \Pi_m(\tau)\:.
\label{ruspm}
\end{equation}
This result immediately follows by a two-fold application of the L\"uders-von-Neumann rule \cite{Luders,vN}.  
With the requirement of a normalized selective post-measurement state (\ref{rPhip}), the work pdf follows from the operation acting on the initial system density matrix as
\begin{equation}
p^{\rm{pem}}_\Lambda(w) = \Tr \Phi_w(\rho(0)) \:.
\label{ppm}
\end{equation}
Hence, the operation $\Phi_w$ can be considered as a complete characterization of the work measurement. It is a linear map on Hilbert space operators with  finite trace that preserves the positivity, i.e., $\rho \geq 0$ implies $\Phi_w(\rho) \geq 0$ and is contractive, i.e.,  $\Tr \Phi_w(\rho) \leq \Tr \rho$ for all $\rho \geq 0$. 
Also more general measurement strategies than the projective one are characterized by operations with  these mathematical properties of being linear, positivity preserving and contractive maps on the space of linear Hilbert space operators with finite trace.  

Coming back to projective measurements we finally want to note the Crooks relation \cite{Tasaki,Crooks,TH,time}. It connects the pdfs $p_\Lambda(w)$ and $p_{\bar{\Lambda}}(w)$ for a protocol $\Lambda =\{\lambda(t)|0\leq t \leq \tau\}$ and the time reversed protocol $\bar{\Lambda} = \{\lambda(\tau-t)|0 \leq t \leq \tau \}$\cite{timereversal}:
\begin{equation}
p^{\rm{pem}}_\Lambda(w) = e^{-\beta (\Delta F -w)} p^{\text{pem}}_{\bar{\Lambda}}(-w)\:.
\label{CR}
\end{equation}
This requires that the initial density matrices of the forward and the backward processes 
 must have Boltzmann weights  as diagonal elements in the energy eigenbasis , i.e. $\Tr \Pi_n(t) \rho(t) = e^{-\beta e_n(t)} d_n(t) Z^{-1}(t)$, where $d_n(t) = \Tr \Pi_n(t)$ is the degeneracy of the energy $e_n(t)$ and $Z(t)= \sum_n e^{- \beta e_n(t)} d_n(t)$ the partition function for $t=0,\tau$.

The free energy difference $\Delta F$ is defined in terms of the partition functions of the initial states of the forward and the backward processes, $e^{-\beta \Delta F} = Z(\tau)/Z(0)$. We note that most often, canonical density matrices are considered as initial states of the forward and backward processes. We point out, however that for the Crooks relation to hold, {\it only the diagonal elements} of these density matrices need to have the form of Boltzmann factors whereas the non-diagonal elements may be arbitrary, as long the density matrix is positive.

\section{Work from two Gaussian energy measurements}
\label{twoenergy}
Instead of the projective energy measurements one may employ generalized measurements of energy for determining work. As demonstrated by Venkatesh {\it et al.} \cite{VWT14}
work distributions obtained by means of generalized measurements typically do not satisfy fluctuation relations. However, for a whole class of generalized energy measurements, in particular for Gaussian measurements, modified fluctuation theorems exist for which the modifications are solely determined by the properties of the measurement apparatus and hence are independent of the force protocol \cite{WVT}. We here choose a less formal approach compared to \cite{WVT} by adapting von Neumann's model \cite{vN} to energy measurements. 
\subsection{Gaussian energy measurement}
In von Neumann's model a measurement apparatus, here called the ``pointer'', is coupled to the system during a short time $\tau_p$. Within this time the value of the observable to be measured is transcribed onto the scale of the pointer while ideally $\tau_p$ is small enough that the unitary dynamics of the system is negligible \cite{fn1}. The information transfer is formally achieved by a unitary operator
$V_t=e^{-i \kappa H\big (\lambda(t)\big) P/\hbar}$ which when acting on the state of the pointer shifts the pointer position  by an amount proportional to the $H\big (\lambda(t) \big )$.  Here, $P$ is the momentum operator which is canonically conjugate to the position $X=\int dx\:x \mathbbm{Q}_x$ of the pointer where $\mathbbm{Q}_x =|x \rangle \langle x|$ is the projection operator on the eigenstate $|x\rangle$ with eigenvalue $x$. The unitary operator $V_t$ describes the time evolution of system and pointer governed within the time span $\tau_p$ by the interaction Hamiltonian $H_{SP}(t) = -g H\big (\lambda(t) \big ) P$ where $g$ describes the coupling strength. The effective interaction strength $\kappa$ between system and pointer  therefore is given by $\kappa = g \tau_p$.   
System and pointer are supposed to be uncorrelated before the interaction has taken place. Then the total density matrix is given by the direct product $\rho \otimes \sigma$ of the system density matrix $\rho$ and the pointer density matrix $\sigma$ immediately before the interaction. 
The system pointer interaction is followed by a projective measurement of the pointer position $x$. The state of the system immediately after the measurement
can be expressed as
\begin{equation}
\rho_x = \phi^{(t)}_x(\rho)/p^{(1)}(x)\:,
\label{rxr}
\end{equation}
where the operation $\phi^{(t)}_x$ is composed of the interaction and the projective measurement, yielding 
\begin{equation}
\begin{split}        
\phi^{(t)}_x (\rho)  &= \Tr_P \mathbbm{Q}_x V_t \rho \otimes \sigma V^\dagger_t\\
 &= \sum_{n,n'} \sigma(x-\kappa e_n(t), x- \kappa e_{n'}(t) ) \Pi_n(t) \rho \Pi(t)_{n'}\:.
\label{phix}
\end{split}
\end{equation}
Here, $\Tr_P$ denotes the trace over the Hilbert space of the pointer, $\sigma(x,y) = \langle x|\sigma |y \rangle$ is the density matrix of the pointer in the position representation, and $p^{(1)}(x)$ gives the probability of observing the pointer in the eigenstate $|x\rangle$. It is determined by the normalization of the post-measurement state $\rho_x$ and therefore becomes
\begin{equation}
\begin{split}
p^{(1)}(x) &= \Tr \phi^{(t)}_x(\rho)\\
&= \sum_n \sigma(x\!-\!\kappa e_n(t), x\!- \!\kappa e_n(t) ) \Tr \Pi_n(t) \rho \:.
\label{p1}
\end{split}
\end{equation} 
We now assume that the initial state of the pointer is given by a Gaussian density matrix with vanishing averages of position and momentum, i.e., $\langle X \rangle =0$ and  $\langle P \rangle =0$. The density matrix $\sigma$ is then uniquely determined by the second moments   $\langle X^2 \rangle$, $\langle XP \rangle$ and $\langle P^2 \rangle$. Note that the positivity of the density matrix is equivalently expressed by the inequality $\langle X^2 \rangle \langle P^2 \rangle \geq |\langle XP \rangle|^2= (\hbar^2 + \langle \{X,P\} \rangle^2)/4$ where $\{X,P\} = XP+PX$ denotes the anti-commutator. In position representation a Gaussian density matrix is given by \cite{T}
\begin{equation}
\sigma(x,y) = \frac{1}{\sqrt{2 \pi \langle X^2 \rangle}} e^{-\mu(x,y)}\:,
\label{sigmaG}
\end{equation}    
where $\mu(x,y)$ is a quadratic form in the position variables $x$ and $y$ defined as
\begin{equation}
\begin{split}
\mu(x,y)&=\frac{1}{2 \hbar^2 \langle X^2 \rangle } \left \{ \langle P^2 \rangle \langle X^2 \rangle (x-y)^2 \right .\\
&\quad \left . - \left [ \langle XP \rangle x - \langle PX \rangle y \right ]^2 \right \}\\
&= \frac{\langle P^2 \rangle}{2 \hbar^2} \left [ x-y \right ]^2 + \frac{1}{2\langle X^2 \rangle }\\
&\quad\times \left [\frac{1}{2} (x+y) - i \frac{\langle \{ X,P \}\rangle}{2 \hbar} (x-y) \right ]^2\:. 
\label{mu}
\end{split}
\end{equation}
Due to the vanishing average pointer position, $\langle X \rangle =0$, the pointer position $x$ presents an unbiased estimate of the system energy having the value $E=x/\kappa$. Hence, the pdf $p(E) = \kappa p^{(1)}(\kappa E)$ to find the system at energy $E$ is given by
\begin{equation}
p(E)= \sum_n \frac{1}{2 \pi \sigma^2_e} e^{-\frac{1}{2 \sigma^2_e}(E-e_n(t))^2}
\Tr \Pi_n(t) \rho\:,
\label{pE}
\end{equation}
where $\sigma^2_e$ is defined by
\begin{equation}
\sigma^2_e = \frac{\langle X^2 \rangle}{\kappa^2}\:.
\label{se}
\end{equation}
For a state with sharp energy $e_k$, i.e. $\Pi_k(t)\rho =\rho$ one obtains from (\ref{pE}) with $\Tr \Pi_n(t) \rho = \delta_{n,k}$ the conditional probability
\begin{equation}
p(E|k) = \frac{1}{2 \pi \sigma^2_e} e^{-(E-e_k(t))^2/(2 \sigma^2_e)}\:.
\label{pEk}
\end{equation}
Hence, $\sigma^2_e$ specifies the inaccuracy of the energy measurement.

Even though the energy pdf (\ref{pE}) agrees with the one obtained in \cite{WVT} the operation $\Phi^{(t)}_E = \kappa \phi^{(t)}_{\kappa E}$ characterizing the measurement generally differs from the form assumed in \cite{WVT}. The latter is defined as $\Phi^{M(t)}_E(\rho) = M_E(t) \rho M^\dagger_E(t)$  in terms of a single  Gaussian Kraus operator $M_E(t) = (2 \pi \sigma^2_e)^{-1/4} e^{-(E-H\big (\lambda(t)\big ))^2/(4 \sigma^2_e)}$. One obtains this particular form from (\ref{phix}) only if the initial pointer density matrix corresponds to a pure Gaussian state in position representation, given by 
\begin{equation}
\sigma(x,y) = \psi_{\text{Gauss}}(x) \psi_{\text{Gauss}} (y)\:,
\label{spure}
\end{equation}
 where 
\begin{equation}
\psi_{\text{Gauss}}(x) = (2 \pi \langle X^2 \rangle)^{-1/4} e^{-x^2/(4 \langle X^2 \rangle)}\:. 
\label{ps}
\end{equation}
In this case 
$\langle X^2 \rangle \langle P^2 \rangle = \hbar^2/4$ and consequently $\langle \{ X,P\} \rangle =0$ holds.

For a general Gaussian initial state specified by (\ref{sigmaG}) and (\ref{mu}) the operation $\Phi^{(t)}_E$ becomes
\begin{equation}
\begin{split}
\Phi^{(t)}_E(\rho)& = \frac{1}{\sqrt{2 \pi \sigma^2_e}} \sum_{n,n'} e^{-\frac{1}{2\sigma^2_{nd}}(e_n(t)-e_{n'}(t))^2}\\
&\quad \times e^{-\frac{1}{2\sigma_e^2} \left [ E -\frac{1}{2}(e_n(t)+e_{n'}(t)) + i \frac{\langle \{X,P \} \rangle}{2 \hbar}(e_n(t) - e_{n'}) \right ]^2 }\\
&\quad \times \Pi_n(t) \rho \Pi_{n'}(t)\:,
\label{PhiG1}
\end{split}
\end{equation}
where 
\begin{equation}
\sigma^2_{nd} = \frac{\hbar^2}{\kappa^2 \langle P^2 \rangle}
\label{snd}
\end{equation}
controls the contribution of non-diagonal elements on the post-measurement state. 
For comparison, the operation $\Phi^{M(t)}_E$ as defined above in terms of a single Gaussian Kraus operator $M_E(t)$ becomes
\begin{equation}
\begin{split}
\Phi^{M(t)}_E (\rho)& =  \frac{1}{\sqrt{2 \pi \sigma^2_e}} \sum_{n,n'} e^{-\frac{1}{8 \sigma^2_e}(e_n(t)-e_{n'}(t))^2}\\
&\quad \times e^{-\frac{1}{2 \sigma^2_e}\left[E-(e_n(t)+e_{n'}(t))/2 \right] }\\
&\quad \times \Pi_n(t) \rho \Pi_{n'}(t)\:.
\label{phiME}
\end{split}
\end{equation}

The subsequent action of $\Phi^{(0)}_E$ followed by the unitary time evolution $U_\Lambda$ and the final operation $\Phi^{(\tau)}_{E'}$ results in an operation characterizing  the joint occurrence of the energies $E$ and $E'$ at the beginning and the end of the force protocol. It is used in the following subsection for characterizing a work measurement composed of two Gaussian energy measurements.

\subsection{Work distribution}
The occurrence  of energies $E$ and $E'$ by two Gaussian measurements separated by the time span of the unitary time evolution under the force protocol 
is characterized by the composed operation $\Phi_{E',E}$ which  acts on a system density matrix $\rho$ as
\begin{equation}
\Phi_{E',E}(\rho) = \Phi^{(\tau)}_{E'}\big (U_\Lambda \Phi^{(0)}_E(\rho) U^\dagger\big )\:.
\label{PhiEE}
\end{equation}
Considering only the resulting work $w=E'-E$ one introduces the according operation $\Phi^{(2)}_w \equiv \int dE \: \Phi_{E+w,E}$ \cite{2} which becomes 
\begin{widetext}
\begin{equation}
\begin{split}
\Phi^{(2)}_w(\rho) &= \frac{1}{\sqrt{4  \pi \sigma^2_e}} \sum_{m,m',n,n'} 
e^{-\frac{1}{2 \sigma^2_{nd}} \left [ (e_m(\tau) -e_{m'}(\tau))^2 + (e_n(0) -e_{n'(0)})^2 \right ] }\\
&\quad \times e^{-\frac{1}{4 \sigma^2_e} \left [w- (w_{m,n} +w_{m',n'})/2 + i \frac{\langle \{ X,P \}\rangle}{2 \hbar} (w_{m,n} - w_{m'n'}) \right ]^2 }
\Pi_m(\tau) U_\lambda \Pi_n(0) \rho \Pi_{n'}(0) U^\dagger_\Lambda \Pi_{m'}(\tau)\:,
\label{Phiw2}
\end{split}
\end{equation}
\end{widetext}
where $w_{m,n} = e_m(\tau) -e_n(0)$ would be a possible work value if the energies were measured projectively.
The non-selective post-measurements state $\rho^{(2)}(\tau^+)$ is given by the integral of $\Phi^{(2)}_w(\rho)$ over $w$ according to (\ref{rhous}) yielding 
\begin{equation}
\begin{split}
\rho^{(2)}(\tau^+)& = \sum_{m,m',n,n'} 
e^{-\frac{1}{2 \sigma^2_{nd}} \left [ (e_m(\tau) -e_{m'}(\tau))^2 + (e_n(0) -e_{n'(0)})^2 \right ] }\\
&\quad \times \Pi_m(\tau) U_\lambda \Pi_n(0) \rho \Pi_{n'}(0) U^\dagger_\Lambda \Pi_{m'}(\tau)\:,
\end{split}
\label{run2}
\end{equation}
The pdf $p^{(2)}(w)$ to find the work $w$ with two Gaussian energy measurements 
is given by the trace of (\ref{Phiw2}) giving
\begin{widetext}
\begin{equation}
\begin{split}
p^{(2)}_\Lambda(w)& = \Tr \Phi_w(\rho(0)) \\
& =\frac{1}{\sqrt{4 \pi \sigma_e^2}}
\sum_{m,n,n'} e^{-\frac{1}{2 \sigma^2_{\text{nd}}}\left [e_n(0)-e_{n'}(0)\right ]^2}  e^{-\frac{1}{4 \sigma^2_e } \left [ w - e_m(\tau) + \frac{1}{2} (e_n(0) + e_{n'}(0)) - i \frac{\langle \{ X,P \} \rangle}{2 \hbar} (e_n(0) - e_{n'}(0)) \right ]^2}p_\Lambda(m,n,n')\:,
\label{p2ew}
\end{split}
\end{equation}
\end{widetext}
where 
\begin{equation}
p_\Lambda(m,n,n') =\Tr \Pi_m(\tau) U_\Lambda \Pi_n(0) \rho(0) \Pi_{n'}(0) U^\dagger_\Lambda\:.
\label{pmnn}
\end{equation}
Note that the diagonal elements of $p_\Lambda(m,n,n')$ with $n=n'$ coincide with the joint probability $p_\Lambda(m,n)$ obtained by two projective energy measurements, see (\ref{pmn}). 
In analogy to (\ref{rPhip}) the selective post-measurement state is given by 
$\rho^{(2)}_w(\tau^+) = \Phi^{(2)}_w(\rho(0))/ p^{(2)}_\Lambda(w)$.
A detailed discussion of the results (\ref{Phiw2}--\ref{p2ew}) is presented in the context of the similar expressions for the work meter discussed next.

\section{A work meter}\label{meter}
Based on a recent idea put forward by Roncaglia, Cerisola and Paz \cite{RCP}, recently, De Chiara, Roncaglia and Paz \cite{dCRP} suggested a method to determine work by means of a single generalized measurement. 
For the sake of completeness we 
describe the principle of operation of such a work meter following Ref. \cite{dCRP} and discuss its functioning in comparison with  two Gaussian energy measurements.  Based thereon, the limiting cases of most accurate and maximally imprecise measurements are investigated as well.
\subsection{Remindful pointer}
The work meter consists of an auxiliary system with a pointer. This pointer interacts twice with the system. The first interaction takes place immediately before the force protocol sets in. It leads to a shift of the pointer proportional to the momentary energy of the system. As for a generalized energy measurement described above the interaction is given by a unitary transformation $V^\dagger_0=e^{i \kappa H\big (\lambda(0) \big ) P/\hbar}$ transferring the negative value of the system energy to the pointer. After the first interaction has ended the position of the pointer is not read out but rather left unchanged until the force protocol has ended.  Then, a second interaction between system and the pointer moves the pointer by an amount that is proportional to the final energy. Accordingly, the interaction is described by the unitary operator $V_\tau = e^{-i \kappa H\big (\lambda(\tau) \big ) P/\hbar}$.
Only after the completion of this interaction, the pointer position is measured projectively. The whole process consisting of the two unitary pointer system interactions, interrupted by the unitary system evolution under the force protocol and the final position measurement can be expressed by a total operation $\phi_x$ characterizing the non-normalized post-measurement state of the system
\begin{equation}
\begin{split}
\phi_x(\rho(0)) &= \Tr_P \mathbbm{Q}_x V_\tau U_\Lambda V^\dagger_0 \rho(0) \otimes \sigma V_0 U^\dagger_\Lambda V^\dagger_\tau\\
&=\sum_{m,m',n,n'} \sigma(x-\kappa w_{m,n},x-\kappa w_{m',n'})\\ &\quad \times \Pi_m(\tau) U_\Lambda \Pi_n(0) \rho(0) \Pi_{n'}(0) U^\dagger_\Lambda \Pi_{m'}(\tau) \:,
\end{split}
\label{phix2}
\end{equation}
As for the generalized energy measurement, $\sigma$ denotes the  density matrix
of the pointer before the first interaction has taken place. Using for $\sigma$ the same Gaussian form (\ref{sigmaG}), (\ref{mu}) together with the pointer calibration $x = \kappa w$, the respective operation  becomes
\begin{widetext}
\begin{equation}
\begin{split}
\Phi_w(\rho(0)) &= \frac{1}{\sqrt{2 \pi \sigma^2_e}}\sum_{m,m',n,n'} e^{-\frac{1}{2\sigma^2_{nd}}(w_{m,n}-w_{m',n'})^2}
e^{-\frac{1}{2 \sigma^2_e}\left [ w - \frac{1}{2}(w_{m,n} +w_{m',n'}) +i\frac{\langle \{ X,P \} \rangle}{2 \hbar} (w_{m,n} -w_{m',n'}) \right ]^2}\\
& \quad \times 
\Pi_m(\tau) U_\Lambda \Pi_n(0) \rho(0) \Pi_{n'}(0) U^\dagger_\Lambda \Pi_{m'}(\tau)\:.
\label{Phiw}
\end{split} 
\end{equation}
\end{widetext}
This expression is similar in its structure to the operation (\ref{Phiw2}) but contains two differences. As a first difference one notices the way how the non-diagonal terms with $n\neq n'$ or $m \neq m'$ are suppressed by the first exponential terms in the sums on the right sides of (\ref{Phiw2}) and (\ref{Phiw}). The exponent is proportional to $(e_m(\tau) - e_{m'}(\tau))^2 + (e_n(0) - e_{n'}(0))^2$ for two energy measurements and $(w_{m,n} -w_{m',n'})^2$ for the work-meter. The second difference stems from the fact that, in the two measurement approach, each energy measurement contributes to the variance of the work and therefore the width of the second exponential factor is twice as large for the work-meter.    

The non-selective post-measurement state $\rho^{\text{wm}}(\tau^+)$ imposed by the work meter (wm) is given by
\begin{equation}
\begin{split}
\rho^{\text{wm}}(\tau^+) &= \sum_{m,m',n,n'} e^{-\frac{1}{2\sigma^2_{nd}}(w_{m,n}-w_{m',n'})^2} \\
& \quad \times \Pi_m(\tau) U_\Lambda \Pi_n(0) \rho(0) \Pi_{n'}(0) U^\dagger_\Lambda \Pi_{m'}(\tau)\:.
\end{split}
\label{runwm}
\end{equation}
It also differs by the weights of the non-diagonal terms from $\rho^{(2)}(\tau^+)$, cf. (\ref{run2}).
The work pdf $p^{\text{wm}}_\Lambda(w) = \Tr \Phi_w\big (\rho(0)\big )$ becomes
\begin{widetext}
\begin{equation}
p^{\text{wm}}_\Lambda(w) = \sum_{m, n,n'} \frac{1}{\sqrt{2 \pi \sigma_e^2 }}
e^{-\frac{1}{2 \sigma^2_{\text{nd}}}\left [e_n(0) - e_{n'}(0) \right ]^2}
e^{-\frac{1}{2 \sigma^2_e} \left [w-e_m(\tau) + \frac{1}{2}(e_n(0) +e_{n'}(0)) - i \frac{\langle \{ X,P \}\rangle}{2 \hbar}(e_n(0)-e_{n'}(0)) \right ]^2} p_\Lambda(m,n,n')\:,
\label{pwp}
\end{equation}
\end{widetext}
Due to the trace the non-diagonal terms $m\neq m'$, which add to the operation $\Phi_w$, do not enter the work pdf. Those non-diagonal terms with $n \neq n'$ contribute with the same exponential weight as for the work pdf resulting from the Gaussian two energy measurement approach. These terms  vanish only for an initially stationary state, i.e. if $[H\big (\lambda(0)\big),\rho(0)] =0$.
Then all non-diagonal elements $p_\Lambda(m,n,n')$ with $n \neq n'$ vanish and the work probability reduces to the following expression:  
\begin{equation}
p^{\rm{d}}_\Lambda(w)=\sum_{m,n} \frac{1}{\sqrt{2 \pi \sigma^2_e}}
e^{-\frac{1}{2 \sigma^2_e} \left [w-e_m(\tau) + e_n(0)  \right ]^2} p_\lambda(m,n,n)\:.
\label{pwd}
\end{equation}
Here, the superscript d stands for ``diagonal'' indicating the restriction to initial density matrices that commute with the initial Hamiltonian and hence are diagonal in the respective eigen-basis.
Because, as already mentioned, the diagonal elements $p(m,n,n)$ agree with the joint probabilities (\ref{pmn}), which determine the work pdf (\ref{pw}) for the two projective energy measurement scheme, the work pdf (\ref{pwd}) becomes a smeared variant of the discrete work distribution obtained by projective energy measurements. The amount of broadening is uniformly determined by the variance $\sigma^2_e = \langle X^2 \rangle / \kappa^2 $, see (\ref{se}).  
The work probability (\ref{pwd}) can now be expressed as the convolution of the projective work pdf and a Gaussian which is solely determined by the properties of the pointer yielding
\begin{equation}
p^{\rm{d}}_\Lambda(w) = \int \frac{d w'}{\sqrt{2 \pi \sigma^2_e}} e^{-(w-w')^2/(2 \sigma^2_e)} p^{\rm{pem}}_\Lambda(w')\:.
\label{pp}
\end{equation}
The pointer properties only enter this expression in the combination of $\sigma_e^2$; in particular, the work pdf does not depend on the pointer momentum variance $\langle P^2 \rangle$ nor on the symmetrized pointer position momentum cross-correlation $\langle \{ P,X \} \rangle$.  It is also worth noticing that the Gaussian factor in the convolution is independent of the force protocol. 

For a canonical initial state $\rho(0) = Z^{-1} e^{-\beta H\big (0\big )}$ the following modified Crooks relation exists \cite{WVT}
\begin{equation}
p^{\rm{d}}_\Lambda(w - \frac{1}{2} \sigma_e^2 \beta) = e^{-\beta( \Delta F -w)} p^{\rm{d}}_{\bar{\Lambda}}(-w-\frac{1}{2} \sigma^2_e \beta)\:,
\label{mCR}
\end{equation}
where $p^{\rm{d}}_{\bar{\Lambda}}(w)$ denotes the work pdf for the time-reversed protocol. As in the original Crooks relation (\ref{CR}), the free energy difference $\Delta F$ refers to the two initial equilibrium states of the backward and the forward processes. From the modified Crooks relation one immediately obtains the modified Jarzynski equality \cite{WVT}
\begin{equation}
\langle e^{-\beta w} \rangle^{\rm{d}} =        e^{-\beta \Delta F} e^{\frac{1}{2}\ \beta^2 \sigma^2_e}\:.
\label{JE}
\end{equation}
The correction factor solely depends on the pointer variance, but is independent of the force protocol.
  
For a pure Gaussian position state of the pointer (pGp), specified by (\ref{spure}, \ref{ps}),  the work pdf  (\ref{pwp}) simplifies to read
\begin{equation}
\begin{split}
  p^{\rm{pGp}}_\Lambda(w) &= \sum_{m,n,n'} \frac{1}{\sqrt{2 \pi \sigma^2_e}}
e^{-\frac{1}{8 \sigma^2_e}\left [e_n(0) - e_{n'}(0) \right ]^2}\\
&\quad \times e^{-\frac{1}{2 \sigma^2_e} \left [w-e_m(\tau) + \frac{1}{2}(e_n(0) +e_{n'}(0)) \right ]^2} p_\Lambda(m,n,n')\:.
\label{mpw}
\end{split}
\end{equation}
In the Section \ref{example} this expression will be illustrated with an example.

\subsection{Accurate measurements}
In the limit of large values of the pointer-system coupling strength $\kappa$, both variances $\sigma^2_{\text{nd}}= \hbar^2/(\kappa^2 \langle P^2 \rangle)$ and $\sigma^2_e= \langle X^2 \rangle / \kappa^2$ defined in (\ref{se}) and (\ref{snd}), respectively, become arbitrarily small. This suppresses all non-diagonal contributions from terms with $n \neq n'$ and further renders the range of deviating work values about each discrete energy difference $e_m(\tau) -e_n(0)$ very narrow. In the limit $\kappa \to \infty$ the work pdf of the work-meter  approaches the two-projective energy measurement result (\ref{pw}), i.e.
\begin{equation}
p^{\text{wm}}_\Lambda(w) \xrightarrow{\kappa \to \infty} p^{\text{pem}}_\Lambda(w)\:.
\label{pk0p}
\end{equation}  
In order to sufficiently suppress all non-diagonal contributions to the sum in (\ref{pwp}), the variance $\sigma^2_{\text{nd}}$ needs to be substantially smaller than the smallest squared eigenenergy difference of the initial Hamiltonian $H\big (\lambda(0)\big )$, i.e.,   
\begin{equation}
2 \sigma^2_{\text{nd}} \ll \min_{\substack{n,n'\\n \neq n'}} \big (e_n(0) -e_{n'}(0)\big )^2\:.
\label{sndee}
\end{equation}
Once the non-diagonal elements are sufficiently suppressed the variance $\sigma^2_e$ must be chosen smaller than the minimum work difference in order to achieve their unambiguous resolution. Hence, the following inequality must be satisfied
\begin{equation}
2 \sigma^2_e \ll \min_{\substack{m,n,m',n'\\ m \neq m' \; \text{or } n \neq n'}} \big ( e_m(\tau) - e_n(0) - e_{m'}(\tau) + e_{n'}(0) \big )^2\:.
\label{swww}  
\end{equation}
Additionally, as mentioned above, the inequality $(1+\langle\{X,P\} \rangle^2/\hbar^2)\sigma^2_{\text{nd}} \leq 4 \sigma^2_e$ must be fulfilled.
An illustrative example is presented in section \ref{example}.    
\subsection{Imprecise measurements}
The opposite limit of maximally imprecise measurements  
is reached for a vanishingly small measurement strength $\kappa$. One expects that in this limit the back-action of the measurement apparatus on the system will be strongly suppressed, however, most likely at the cost of a large measurement error.  Before confirming this qualitative picture on the basis of the work pdf  (\ref{pwp}) we first consider the mean value of the work $\langle w \rangle = \int dw\:w\: p_\Lambda(w)$ in this particular limit.
Using (\ref{pwp}) we obtain for the average work
\begin{widetext}
\begin{equation}
\langle w \rangle =  \sum_{\substack{m \\ n,n'}} e^{-\frac{1}{2 \sigma^2_{nd}}\big (e_n(0) - e_{n'}(0)\big )^2}
\left [e_m(\tau)- \frac{1}{2} \big (e_n(0)+e_{n'}(0) \big ) + i \frac{\langle \{ X,P \} \rangle }{2 \hbar} \big (e_
n(0) -e_{n'}(0) \big ) \right]
p_\Lambda(m,n,n') \:.
\label{mw}
\end{equation}
Using (\ref{pmnn}) one can further simplify this average by performing the sum on $m$ and splitting the remaining sum into its diagonal and non-diagonal contributions leading to 
\begin{equation}
\begin{split}
\langle w \rangle &= \Tr H\big (\lambda(\tau)\big ) \bar{\rho}(0) - \Tr H\big (\lambda(0)\big ) \rho(0) \\
&\quad 
+ \sum_{n>n'} e^{-\frac{1}{2 \sigma^2_{\text{nd}}}(e_n(0) - e_{n'}(0))^2}  \Tr H\big (\lambda(\tau)\big ) U_\Lambda \big ( \Pi_n(0) \rho(0) \Pi_{n'}(0) + \Pi_{n'}(0) \rho(0) \Pi_n(0) \big ) U^\dagger_\Lambda \:,
\label{aw}
\end{split}
\end{equation}
\end{widetext}
where
\begin{equation}
\bar{\rho}(0) = \sum_n \Pi_n(0) \rho(0) \Pi_n(0)
\label{br0}
\end{equation}
denotes the projection of the initial density matrix onto the eigen-basis of the initial Hamiltonian $H\big(\lambda(0)\big )$. 
The expression (\ref{aw}) for the average work is still exact. The first two terms on the right hand side coincide with the average work obtained from two projective energy measurements; the sum contains contributions from  non-diagonal elements of the initial density matrix. 
Note that the imaginary term on the right hand side of (\ref{mw}), which is caused by a finite symmetrized pointer position momentum correlation, does not contribute to the average.

In the imprecise limit $\kappa \to 0$ the exponential term under the sum on the right hand side approaches unity. Then, the sums in (\ref{aw}) can be performed to yield the work as the difference of the energy expectation values at the end and the beginning of the force protocol, leading to 
\begin{equation}
\langle w \rangle = \Tr_S H\big (\lambda(\tau)\big ) U_\Lambda \rho(0) U^\dagger_\Lambda - \Tr_S H\big (\lambda(0)\big ) \rho(0)\:.
\label{wHH}
\end{equation}
As expected, this result does not contain any back-action of the first energy measurement and coincides with the untouched work average (\ref{varw}). Hence, one might be tempted to expect that the imprecise limit could provide a physical access to untouched work.

In order to study this limit in more detail, we consider the characteristic  function of work \cite{TLH}, which is defined as the Fourier transform of the work pdf.
Hence it becomes
\begin{equation}
\begin{split}
G_\Lambda(u) &= \int dw e^{i u w} p_\Lambda(w) \\
&= \sum_{\substack{m \\ n,n'}} p_\Lambda(m,n,n') \chi_{m,n,n'}(u)\:,
\label{G}
\end{split}
\end{equation}
where $\chi_{m,n,n'}(u)$ is determined by the Fourier transformed pointer density matrix 
\begin{equation}
\begin{split}
\chi_{m,n,n'}(u) &= \int dx e^{i u x/\kappa} \sigma(x-\kappa w_{m,n},x-\kappa w_{m,n'}) \\
& = e^{-\frac{1}{2 \sigma^2_{\text{nd}}} (e_n(0)-e_{n'}(0))^2} e^{-\frac{1}{2} \sigma^2_e u^2} e^{i u c_{m,n,n'}} 
\label{chi}
\end{split}
\end{equation}
with 
\begin{equation}
c_{m,n,n'} = e_m(\tau) -\alpha e_n(0)  - \alpha^* e_{n'}(0) 
\label{cmnn}
\end{equation}
and 
\begin{equation}
\alpha= \frac{1}{2} (1 - \frac{i}{\hbar} \langle \{X,P \} \rangle )\:.
\label{alpha}
\end{equation}
The expression (\ref{G}) with (\ref{chi}) and (\ref{cmnn})  is still exact. It can be further evaluated in the imprecise limit in which the first, u-independent exponential factor on the right hand side of (\ref{chi}) can be replaced by unity. The triple sum in (\ref{G}) can then be performed to yield the characteristic function in the imprecise limit, $G^{\text{il}}_\Lambda(u)$, reading  
\begin{equation}
G^{\text{il}}_\Lambda(u) = e^{-\frac{1}{2 } \sigma^2_e u^2} g_\Lambda(u) \:,
\label{Gilu}
\end{equation}
where
\begin{equation}
g_\Lambda(u) = \Tr e^{iu H^H\big (\lambda(\tau)\big )}  e^{-i \alpha u H\big (\lambda(0)\big )} \rho(0)e^{-i \alpha^* u H\big (\lambda(0)\big )}\:.
\label{gLu}
\end{equation}
Further,  $H^H\big ( \lambda(\tau) \big) \equiv U^\dagger_\Lambda H\big (\lambda(\tau) \big ) U_\Lambda$ denotes the final Hamiltonian in the Heisenberg picture.
The respective work pdf $p^{\text{il}}(w)$ results as a sum of wide Gaussians, i.e.,  
\begin{equation}
\begin{split}
p^{\text{il}}_\Lambda(w) &= \sum_{\substack{m \\ n,n'}} \frac{1}{\sqrt{2 \pi \sigma^2_e}} e^{-\frac{1}{2 \sigma^2_e}\left [w-e_m(\tau)+\alpha e_n(0) +\alpha^* e_{n'}(0) \right ]^2}\\
&\quad \times p_\Lambda(m,n,n')
 \label{pilw}
\end{split}
\end{equation}
Provided the inverse Fourier transform of $g_\Lambda(u)$ exists the work pdf can also be expressed as a convolution with a Gaussian of width $\sigma_e$, i.e.
\begin{equation}
p^{\text{il}}_\Lambda(w) = \int \frac{dw'}{\sqrt{2 \pi \sigma^2_e}} e^{-\frac{1}{2 \sigma^2_e}(w-w')^2} q_\Lambda(w')\:,
\label{pilwc}
\end{equation}
where
\begin{equation}
q_\Lambda(w) = \int \frac{du}{2 \pi} e^{-iuw} g_\Lambda(u)\:.
\label{qg}
\end{equation}
is the inverse Fourier transform of $g_\Lambda(w)$.

Before we further discuss the condition of existence and the properties of $q_\Lambda(w)$ we note that the result given in the first line of (\ref{pilw}) can also be directly obtained from the work pdf (\ref{pwp}) by replacing the first exponential on the right hand side of (\ref{pwp}) by unity. In this way all non-diagonal elements contribute equally.

The expression (\ref{gLu}) coincides with the characteristic function of work obtained by two projective energy measurements \cite{TLH} only if the initial density matrix $\rho(0)$ commutes with the Hamiltonian $H\big (\lambda(0)\big )$ at the beginning of the protocol. Then the two exponents containing $H\big (\lambda(0) \big)$ can be combined into a single term such that $g_\Lambda(u) = G^{\text{pem}}(u) \equiv \Tr e^{i u H^H\big ( \lambda(\tau) \big )} e^{-iu H\big ( \lambda(0) \big )} \rho(0)$ and $q_\Lambda(w) = p^{\text{pem}}_\Lambda(w)$ hold.  In general, though, the inverse Fourier transform of $g_\Lambda(u)$ leads to a divergent result due to the presence of the imaginary part of the coefficient $\alpha$.  Only if  the symmetrized pointer position momentum cross-correlation vanishes, $\langle \{X,P \}\rangle =0$,   $\alpha$ is real and $q_\Lambda(w)$ becomes a sum of the projective work pdf $p^{\text{pem}}_\Lambda(w)$  and further delta-contributions at work values that correspond to the differences $e_m(\tau) -\big (e_n(0) + e_{n'}(0)\big )/2$ between the final energy eigenvalues and the arithmetic averages of all pairs of initial eigenvalues. Hence, for $\langle \{ X,P \} \rangle =0$, the inverse Fourier transform of the trace expression becomes
\begin{equation}
\begin{split}
q_\Lambda(w) &= p^{\text{pem}}_\Lambda(w)\\
&\quad + \sum_{\substack{m,n,n'\\n\neq n'}} \delta \left (w- e_m(\tau) + \frac{1}{2} (e_n(0) + e_{n'}(0)) \right )\\
& \quad \times p_\Lambda(m,n,n')\:.
\label{qp}
\end{split}
\end{equation}
The integral of $q_\Lambda(w)$ over all possible work values is unity as 
follows from (\ref{pilw}) with $\int dw q_\Lambda(w) = g_\Lambda(0)=1$. 
Because 
 the projective work pdf $p^{\text{pem}}_\Lambda(w)$ 
itself is normalized the integral over the sum expression on the right hand side must vanish. This implies that either this sum vanishes for all values of $w$ or that it assumes both positive and negative values. The first case happens only if $p_\Lambda(m,n,n') =0$ for all $n \neq n'$. One can show that for this to hold  the initial density matrix must be stationary, i.e.  $[H\big (\lambda(0)\big ),\rho(0)]=0$. We conclude that in all other cases $q_\Lambda(w)$ assumes both positive and negative values and therefore does not qualify as a proper pdf. The positivity of $p^{\text{il}}_\Lambda(w)$ is only guaranteed by the convolution with a sufficiently broad Gaussian which smooths out any negative contributions.

In agreement with the expectation formulated in the beginning of this subsection the non-selective post-measurement state (\ref{runwm}) converges to the state that is reached solely by the force protocol without the inference of the work meter, i.e. one obtains
\begin{equation}
\rho^{\text{wm}}(\tau^+) \xrightarrow{ \kappa \to 0} U_\Lambda \rho(0) U^\dagger_\Lambda\:.
\label{upmk0}
\end{equation}

Provided $q_\Lambda(w)$ exists, the convolution in the second line of (\ref{pilw}) can be evaluated in the limit $\sigma_e \to \infty$ yielding
\begin{equation}
p^{\text{il}}_\Lambda(w) \approx \frac{1}{2 \pi \sigma^2_e} e^{-\frac{1}{2\sigma^2_e}(w - \prec w \succ)^2}\:.
\label{pas}
\end{equation}
We conclude that the imprecise limit suppresses any influence of the interactions of the work measurement device and in this sense yields the untouched work but on the other hand does not carry any system-specific information on the work beyond its average.  

Finally, we note that the limiting form (\ref{pilw})   presents a valid approximation of the work pdf (\ref{pwp}) if 
\begin{equation}
2 \sigma^2_{nd} \gg (e_{n^*} - e_N)^2\:, 
\label{snde}
\end{equation}
where both $n^*$ and $N$ are elements of the set $\mathcal{N}_\epsilon$ that exhaust the total probability obtained from the 
diagonal elements $p_n = \Tr \Pi_n(0) \rho(0)$ up to a negligibly small fraction $\varepsilon$, such that $\sum_{n \in \mathcal{N}_\epsilon} p_n =1- \varepsilon $.   Here, $e_{n^*} = \min_{n \in \mathcal{N}_\epsilon} e_n(0)  $ and $e_N = \max_{n \in \mathcal{N}_\epsilon} e_n(0)$ are the minimum and maximum energies, respectively, out of this set of states contributing to the energy distribution in the initial state. The criterion (\ref{snde}) is obtained from the fact that the non-diagonal elements of $p_\Lambda(m,n,n')$ are proportional to $\rho_{n,n'} =\langle n;0|\rho(0) |n';0\rangle$, cf. (\ref{pmnn}), \cite{nondegenerate}. Due to the positivity of the density matrix $\rho(0)$, the absolute values of these non-diagonal elements of the density matrix are dominated by the diagonal elements as $|\rho_{n,n'}|^2 \leq p_n p_{n'}$. 
   
\section{Work distributions of a quenched two level system}
\label{example}    
\begin{figure}
\includegraphics[width=8cm]{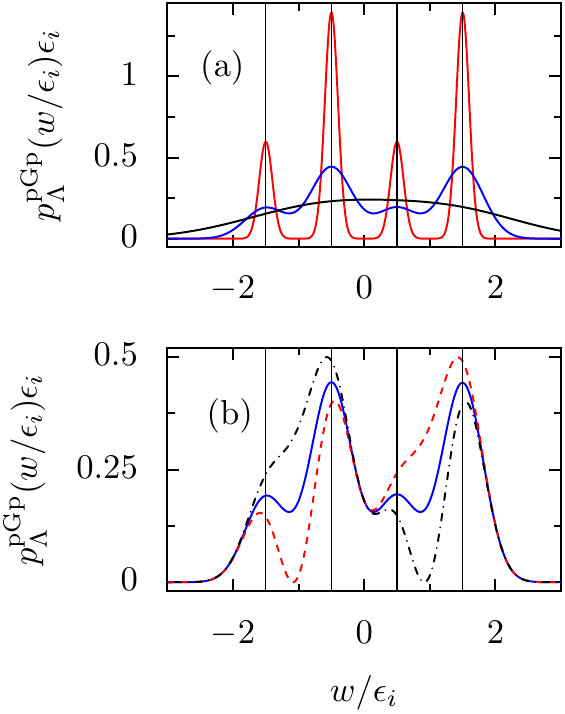}
\caption{The pdf of work supplied on a two-level system by  a sudden change  of the Hamiltonian from $H_i$ to $H_f$ given by (\ref{Hi}) and (\ref{Hf}), respectively, is displayed as it is generated by a work meter with a pure Gaussian state in position representation. The level separation of the Hamiltonian after the switch is given by $\epsilon_f = 2 \epsilon_i$. Both panels refer to an initial density matrix (\ref{rho0}) with ground-state probability $p=0.7$. In panel (a) the work pdf is displayed for vanishing non-diagonal elements of the initial density ($q=0$) and different values of the work variance $\sigma^2_e$ which are  $\sigma^2_e = 0.01 \times \epsilon^2_i$ (red), $0.1 \times \epsilon^2_i$ (blue), $\epsilon^2_i$ (black). At the smallest variance, the work pdf consists of clearly separated peaks, which overlap at the intermediate variance and are completely washed out at the large variance. In those cases where peaks can be identified they are always exactly located at the positions of the work values (\ref{w12}) obtained by projective measurements. These values are indicated by the four thin vertical black lines. Panel (b) exemplifies the effect of a non-diagonal initial density matrix at the intermediate variance $\sigma^2_e =0.1 \times \epsilon^2_i$. As a reference, the solid blue line refers to the diagonal case ($q=0$) whereas the red dashed curve corresponds to $q=\sqrt{0.21}$ and the black dash-dotted line to  $q=-\sqrt{0.21}$. The non-diagonal elements of the initial density matrix $\rho(0)$ lead to a substantial distortion of the work pdf. In particular, the positions of the maxima do no longer coincide with those of the projective work values.}
\label{f1}
\end{figure} 
\begin{figure}
\includegraphics[width=8cm]{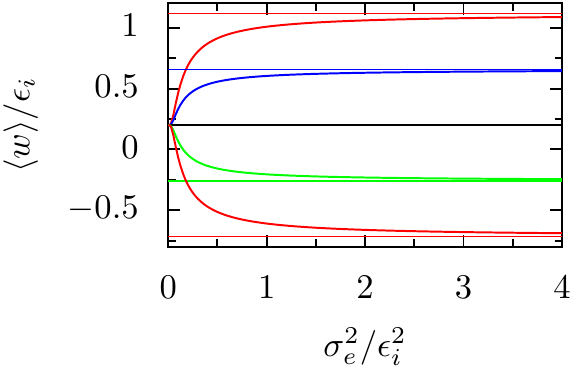}
\caption{  The average work approaches asymptotically the difference of untouched energy mean values $\prec \!w \!\succ$ (\ref{varw}) as a function of the variance $\sigma^2_e$, see  (\ref{mw}). All curves are for a density matrix of the type (\ref{rho0}) with $p=0.7$. The average work is displayed for $q=\sqrt{0.21}$ by the upper red curve, for $q=0.5 \times \sqrt{0.21}$, $0$, $-0.5\times\sqrt{0.21}$ and $-\sqrt{0.21}$ by the blue, green und lower red curves, respectively. The according asymptotic values given by (\ref{mw}) are indicated by thin horizontal lines of respective color.  The final energy is chosen as $\epsilon_f = 2 \times \epsilon_i$. The change in the average work is largest for the two limiting values $q=\pm \sqrt{p(1-p)}=\pm \sqrt{0.21}$ and vanishes for $q=0$. }
\label{f2}
\end{figure} 
\begin{figure}
\includegraphics[width=8cm]{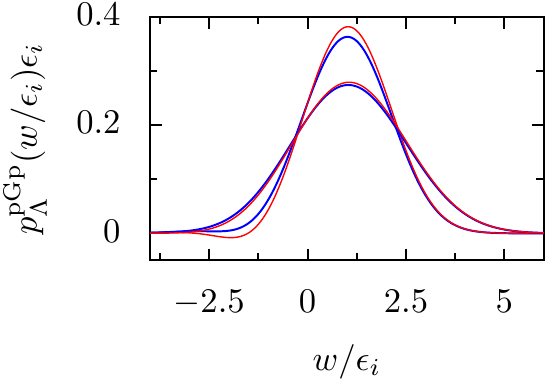}
\caption{The exact work pdf for $\sigma^2_e =\epsilon^2_i$ (narrow blue curve)
and $\sigma^2_i = 2 \times \epsilon^2_i$ (broad blue curve) is compared to the limiting result of the imprecise measurement limit for two values of the variance $\sigma^2_e$ (thin red curves). At the smaller value $\sigma^2_w =\epsilon^2_i$ the Gaussian factor of the convolution integral on the right hand side of (\ref{pilw}) is not yet broad enough to suppress the negative contributions around $w = - 2 \times \epsilon_i$. At the larger variance, the approximate work pdf only insignificantly differs from the exact one. The other parameters, $p=0.7$ and $q= \sqrt{0.21}$, are the same for all curves.}
\label{f3}
\end{figure}

As a simple example we consider a spin 1/2 which undergoes a sudden change of the Hamiltonian from its initial form
\begin{equation}
H_i = \frac{\epsilon_i}{2} \sigma_z
\label{Hi}
\end{equation}
to the final form
\begin{equation}
H_f = \frac{\epsilon_f}{2} \sigma_x\:,
\label{Hf}
\end{equation}
where $\sigma_x$ and $\sigma_z$ are Pauli-spin matrices.
The central quantity $p_\Lambda(m,n,n)$ can then be expressed in terms of the eigenfunctions $|n;i\rangle$ and $|n;f\rangle$ of $H_i$ and $H_f$, respectively as
\begin{equation}
p_\Lambda(m,n,n') = \langle m;f|n;i \rangle \langle n;i|\rho(0)| n',i\rangle \langle n',i| m;f\rangle\:.
\label{pLmnn}
\end{equation}
The indices $m,n,n'$ taking the values 1 and 2 refer to the ground and the exited state, respectively, of each Hamiltonian. The scalar products are readily determined to yield $\langle 1,i|1,f \rangle =-\langle 2,i|1,f \rangle=\langle 1,i|2,f \rangle=\langle 2,i|2,f \rangle= 1/\sqrt{2}$. 

We specify the density matrix $\rho(0)$ in the basis of $H_i$ as
\begin{equation}
\rho(0) = \left ( \begin{array}{cc}
p & q \\
q^*&1-p
\end{array}
\right )
\label{rho0}
\end{equation}
with $0\leq p \leq 1$ and $p(1-p)\geq |q^2|$. 

Choosing for the initial pointer state $\sigma$ a pure Gaussian state as defined in (\ref{spure}, \ref{ps}),
 we obtain
\begin{equation}
\begin{split}
p^{\text{pGp}}_\Lambda(w) &=p^{\text{diag}}_\Lambda(w) +\frac{q+q^*}{2 \sqrt{2 \pi \sigma^2_e}} e^{-\frac{\epsilon^2_i}{8 \sigma_e^2}}\\
& \times \left [e^{-\frac{1}{2 \sigma^2_e}(w-\epsilon_f/2)^2} - e^{-\frac{1}{2 \sigma^2_e} (w+\epsilon_f/2)^2} \right ]\:,
\label{pw2l}
\end{split} 
\end{equation}
where the diagonal elements of the initial density matrix contribute with
\begin{equation}
\begin{split}
p^{\text{diag}}_\Lambda(w) &= \frac{1}{\sqrt{2 \pi \sigma^2_e}} \left \{ \left [e^{-\frac{1}{2 \sigma^2_e}(w-w_{1,1})^2} +e^{-\frac{1}{2 \sigma^2_e}(w-w_{2,1})^2} \right ] p \right . \\
&\quad \left . + \left [e^{-\frac{1}{2 \sigma^2_e}(w-w_{1,2})^2} +e^{-\frac{1}{2 \sigma^2_e}(w-w_{2,2})^2} \right ] (1-p) \right \}\:.
\end{split}
\label{pdiag2l}
\end{equation}
The allowed projective work-values $w_{m,n}$ are given by
\begin{equation}
\begin{split}
w_{1,1} = -\epsilon_f/2 + \epsilon_i/2, \quad & w_{2,1} = \epsilon_f/2 + \epsilon_i/2\\     
w_{1,2} = -\epsilon_f/2 - \epsilon_i/2, \quad & w_{2,2} = \epsilon_f/2 - \epsilon_i/2\:.
\end{split}
\label{w12}
\end{equation}
Figure \ref{f1} displays the work pdf (\ref{pw2l}) for different variances $\sigma^2_e$ and different initial system density matrices. While an increase of $\sigma^2_e$ leads, as expected, to a broadening of the peaks, the presence of non-diagonal terms $q\neq 0$ in the system's density matrix gives rise to further deformations and, in particular, also to shifts of these peaks.
Figure \ref{f2} demonstrates the transition of the average work from the accurate limit yielding the results of projective energy measurements to the limit of maximally imprecise measurements for different values of the non-diagonal elements of the initial system's density matrix. In the latter limit the approach to the average untouched work (\ref{varw}) is confirmed. Finally, in Figure \ref{f3} the asymptotic work pdf (\ref{pilw}) is compared with the full result (\ref{pw2l}) for two different values of the variance $\sigma^2_e$. For the smaller one, a negative contribution resulting from negative parts of $q_\Lambda(w)$ is visible whereas for the larger one 
the Gaussian factor in (\ref{pilw}) is sufficiently broad to suppress any negative part.
As a result, the limiting distribution coincides fairly well with the exact one, even though with $2\sigma^2_{nd} = 16 \epsilon^2_i$ and $(e_{n^*}-e_N)^2 =8 \epsilon^2_i$ the inequality (\ref{snde}) is only marginally satisfied.

\section{Conclusions}\label{con}
We reviewed several attempts to specify work. In particular, we considered three different setups for which we compared the state of the considered system immediately after the completion of selective and non-selective measurements of work as well as the respective probabilities of finding a particular value of work. In the first scenario, work is determined in the standard way by two projective measurements of energy; in the second scenario, also two energy measurements are performed in each of which the system is coupled to a Gaussian pointer state and read out by projective measurements of the pointer state; the third scenario, proposed by De Chiara, Roncaglia and Paz in \cite{dCRP}, the result of the first pointer-system interaction encoding the negative initial energy is stored in the pointer until the force protocol is completed. The final energy is transferred to the same pointer and added to the stored value. Only then the pointer state is read out by a projective measurement.

In all three cases the operations characterizing the selective as well as the non-selective post-measurement states and the distributions of work values are different. While the projective energy measurements select the  elements of the according density matrix that are diagonal with respect to the measured Hamiltonians, non-diagonal elements contribute to the post-measurements states generated by both Gaussian strategies, however with different weights. Moreover, when considering the work pdfs one finds  broadened peaks instead of the sharp lines in the case of projective energy measurements. The widths of these peaks are governed by an effective measurement strength, $\sigma^2_e= \langle X^2 \rangle /\kappa^2$,  which is determined by the variance of the pointer position $\langle X^2 \rangle$, in its initial state and a parameter $\kappa = g \tau_p$ combining  the strength $g$ and duration $\tau$ of the interaction between system and pointer. A second variance-like quantity ($\sigma^2_{\text{nd}} = \hbar^2/(\langle P^2 \rangle \kappa$) governs the influence of non-diagonal terms of the initial system density matrix. These terms not only add to the broadening but also cause shifts of the peaks of the work pdf relative to the possible work values obtained by projective energy measurements.  The two variances are restricted by an inequality which suppresses the influence of non-diagonal elements of the initial density matrix if the broadening of the lines is small. On the other hand it entails huge broadening if the non-diagonal elements are only little suppressed. The disjoint parameter regimes of small $\sigma_e$ and large $\sigma_{nd}$ correspond to the limiting cases of accurate and imprecise measurements. In the accurate limit the work statistics based on projective energy measurements is approached, whereas in the imprecise limit the work statistics is governed by an extremely broad Gaussian distribution with an average of the untouched work $\prec w \succ$ and width $\sigma_e\gg |\!\prec w \succ\!|$. In this limit, the fluctuations are completely determined by the imprecision of the measurement device. Therefore this limit is of minor practical use; to achieve a reliable estimate of the average untouched work an extremely large number of realizations would be required.

Finally, we note that the fluctuation relations of Jarzynski and Crooks remain valid in its modified form only if the system initially stays in a canonical state. In contrast to the projective two energy measurement approach it is not sufficient if only the diagonal elements in the initial energy basis are given by Boltzmann factors but additional finite non-diagonal elements are present.           
     
\acknowledgments
PT thanks the
Foundation for Polish Science (FNP) for providing him with an Alexander von Humboldt Polish Honorary
Research Fellowship.


\begin{thebibliography}{99}
\bibitem{gauge}Note that for non-autonomous Hamiltonian systems not any Hamiltonian rendering the correct equations of motion gives a proper expression for the energy. However, with a convenient gauge transformation it can always be achived that the value of the Hamiltonian corresponds to the instantaneous energy of the system \cite{CHT}.
\bibitem{untouched} We choose the name ``untouched work'' because the first registration of energy is assumed to leave the subsequent dynamics untouched.
\bibitem{fluctuations} For the two-ensemble implementation of untouched work the fluctuations of work are given by the pdf $p^{\text{te}}(w) =\sum_{m,n} \delta(w-e_m(\tau)+e_n(0)) \Tr \Pi_m(\tau) U_\Lambda \rho(0) U^\dagger_\Lambda \Tr \Pi_n(0) \rho(0)$. This expression, though does not have the correct classical limit and therefore cannot be considered as a useful definition.
\bibitem{time} In order that the time-reversed process can be realized, the Hamiltonian of the system must satisfy the time-reversal symmetry relation $H\big (\lambda(t)\big ) = \theta H\big (\epsilon \lambda(t)\big ) \theta^\dagger$ for all times $t \in [0,\tau]$ with $\theta$ being the time-reversal operator and $\epsilon =\pm 1$  the parity of the parameter $\lambda$ under time-reversal.
\bibitem{timereversal} For the sake of simplicity, we consider a parameter $\lambda(t)$ that transforms evenly under time-reversal. For odd parameters, its sign must be changed.
\bibitem{fn1}The assumptions of a short interaction time $\tau_p$ during which the system dynamics is negligible is made for the sake of simplicity. One may allow for larger interaction times during which the system dynamics is no longer negligible \cite{VWT15,Prinz} without affecting the general conclusions.
\bibitem{2} The superscript $2$ is meant to distinguish a work operation based on two Gaussian energy measurements in contrast to the operation characterizing the work meter discussed below, see Section \ref{meter}.
\bibitem{nondegenerate} For the sake of simplicity, we assume that the energy eigenvalues $e_n(0)$ of $H\big (\lambda(0) \big )$ with corresponding eigenstates  $|n;0 \rangle$ are non-degenerate.
\end{thebibliography}
\end{document}